\documentclass[english,12pt,aps,prd,a4paper,preprintnumbers,floatfix,nofootinbib,showpacs,superscriptaddress, notitlepage]{revtex4-1} 
 \pdfoutput=1
\usepackage[usenames,dvipsnames]{color}  
\usepackage{graphicx}
\usepackage{pifont}
\usepackage{caption}
\usepackage{subcaption}
\usepackage{amsmath}
\usepackage{amssymb}
\usepackage[colorlinks=true,citecolor=darkred,urlcolor=darkred, pdfborder={0 0 0}]{hyperref}
\usepackage[normalem]{ulem}

\usepackage[T1]{fontenc}
\usepackage[latin9]{inputenc}
\usepackage{array}
\usepackage{booktabs}
\usepackage{mathrsfs}
\usepackage{multirow}
\usepackage{tabularx}

\definecolor{darkred}{rgb}{0.6,0,0}

\definecolor{linkcolor}{rgb}{0,0,0.5}

\def\gsim{\raise0.3ex\hbox{$\;>$\kern-0.75em\raise-1.1ex\hbox{$\sim\;$}}}
\def\lsim{\raise0.3ex\hbox{$\;<$\kern-0.75em\raise-1.1ex\hbox{$\sim\;$}}}

\def\beqn#1{\begin{equation}\label{#1}}
\def\eeqn{\end{equation}}

\def\beqa#1{\begin{eqnarray}\label{#1}}
\def\eeqa{\end{eqnarray}}

\def\Z2{$\mathcal{Z_2}$}

\newcommand {\ignore}[1]{}

\newcommand{\sm}{{Standard Model }}

\def\cevns{CE$\nu$NS~}
\def\321{$\mathrm{SU(3) \otimes SU(2) \otimes U(1)}$ }

\newcommand{\AddrAHEP}{%
  AHEP Group, Institut de F\'{i}sica Corpuscular --
  CSIC/Universitat de Val\`{e}ncia, Parc Cient\'ific de Paterna.\\
 C/ Catedr\'atico Jos\'e Beltr\'an, 2 E-46980 Paterna (Valencia) - SPAIN}

\newcommand{\AddrMiranda}{%
Departamento de F\'{\i}sica, Centro de Investigaci\'on
  y de Estudios Avanzados del IPN,\\ Apartado Postal 14-740 07000 Mexico,
  Distrito Federal, Mexico}

\begin{document}

\bibliographystyle{unsrt}   

\title{\boldmath \color{BrickRed} Probing new neutral gauge bosons \\ with \cevns and neutrino-electron scattering}

\author{O.G. Miranda}\email{omr@fis.cinvestav.mx}\affiliation{\AddrMiranda}
\author{D.K. Papoulias}\email{dipapou@ific.uv.es}\affiliation{\AddrAHEP}
\author{M. T\'ortola}\email{mariam@ific.uv.es}\affiliation{\AddrAHEP}
\author{J. W. F. Valle}\email{valle@ific.uv.es}\affiliation{\AddrAHEP}

\begin{abstract}
The potential for probing extra neutral gauge boson mediators ($Z^\prime$) from low-energy measurements is comprehensively explored. 
Our study mainly focuses on $Z^\prime$ mediators present in string-inspired $E_6$ models and left-right symmetry. 
We estimate the sensitivities of coherent-elastic neutrino-nucleus scattering (CE$\nu$NS) and neutrino-electron scattering experiments.
Our results indicate that such low-energy high-intensity measurements can provide a valuable probe, complementary to high-energy collider searches and electroweak precision measurements. 
\end{abstract}

\maketitle

\section{Introduction}

Despite its amazing success~\cite{Tanabashi:2018oca} it is well-accepted that the \sm (SM)  can not be the whole truth.
Although the SM seems to capture the most essential features concerning the gauge description of fundamental interactions, it leaves many in the open.
Indeed, many are the theoretical motivations for having an extended gauge structure.
The latter include the desire of incorporating a dynamical seesaw mechanism that can naturally account for small neutrino masses~\cite{Schechter:1980gr,Schechter:1981cv} in such a way that these are linked to the origin of parity violation in the weak interaction~\cite{Mohapatra:1980yp}. Embedability into a simple unified structure at high energies~\cite{GellMann:1980vs} also motivates the existence of new gauge bosons.

Searches for heavy intermediate vector bosons have been extensively performed using high energy accelerators such as the LHC~\cite{Aad:2019fac}.
Their existence could also have important implications for electroweak precision tests~\cite{GonzalezGarcia:1990yq,GonzalezGarcia:1990tr,Langacker:2008yv} and induce
charged lepton flavor violation~\cite{Das:2012ii,Deppisch:2013cya}. 
Building on early work~\cite{Miranda:1997vs,Garces:2011aa} here we will examine the sensitivity of a number of experimental low-energy setups to the existence of heavy electrically neutral intermediate vector bosons $Z^\prime$.
These are expected in theories with gauged B-L~\cite{pati:1974yy,Mohapatra:1980qe}, 
in extended electroweak models predicting the number of families~\cite{Singer:1980sw,Hati:2017aez},
in models with dynamical symmetry breaking~\cite{Akhmedov:1995vm,Akhmedov:1995ip,Hill:2002ap}, 
in string-inspired extensions of the SM \cite{Cvetic:1995rj},
as well as in ambitious ``comprehensive unification'' scenarios with extra dimensions~\cite{Reig:2017nrz}.
It has been shown that a neutral $Z^\prime$ can have masses at the TeV scale in a way consistent with neutrino mass generation as well as gauge coupling unification in $SO(10)$~\cite{Malinsky:2005bi}.
In this work we focus on scenarios where a $Z^\prime$ boson has mass around the few TeV scale. 

The recent discovery of coherent elastic neutrino-nucleus scattering (CE$\nu$NS) by the COHERENT experiment~\cite{Akimov:2017ade} at the Spallation Neutron Source (SNS), has inspired many phenomenological studies addressing the sensitivity of low-energy approaches to new physics (for a review see Ref.~\cite{Papoulias:2019xaw}).
New constraints have been placed on nonstandard and generalized interactions~\cite{Liao:2017uzy,Giunti:2019xpr,Coloma:2019mbs,AristizabalSierra:2018eqm}, nuclear physics parameters~\cite{Cadeddu:2017etk,Papoulias:2019lfi,Canas:2019fjw}, neutrino  electromagnetic properties~\cite{Cadeddu:2018dux,Miranda:2019wdy}, sterile neutrinos~\cite{Miranda:2019skf,Blanco:2019vyp,Berryman:2019nvr} as well as implications for dark matter~\cite{Ge:2017mcq}. 
It has been noted~\cite{Dutta:2015vwa,Dent:2016wcr,Billard:2018jnl,Datta:2018xty,Denton:2018xmq,Farzan:2018gtr,AristizabalSierra:2019ufd,Dutta:2019eml,AristizabalSierra:2019ykk,Arcadi:2019uif} that light mediators may be accessible to \cevns experiments, providing information that is complementary to what can be achieved from high-energy and/or precision measurements. Constraints on the $Z^\prime$ parameters from \cevns have been reported in Refs.~\cite{Liao:2017uzy,Kosmas:2017tsq}. 

In the present paper, we further explore the complementarity of the high-intensity, low-energy approach as a tool to search for new physics.
In the same spirit as Refs.~\cite{Miranda:1997vs,Garces:2011aa}, here we consider the sensitivity of various low-energy experimental setups involving \cevns and neutrino-electron scattering
to the existence of extra neutral gauge bosons arising from well-motivated   left-right (LR) symmetric and $E_6$-based  theories. Regarding SNS neutrinos, in addition to the ``first light'' \cevns  measurement with a CsI[Na] detector, we also explore the new physics potential at the future Ge,  liquid argon (LAr)     and NaI[Tl] detector subsystems of COHERENT~\cite{Akimov:2018ghi}. In addition, we test the corresponding capabilities at various proposed reactor-based \cevns facilities such as CONUS~\cite{Hakenmuller:2019ecb}, CONNIE~\cite{Aguilar-Arevalo:2019jlr}, MINER~\cite{Agnolet:2016zir}, TEXONO~\cite{Wong:2010zzc}, RED100~\cite{Akimov:2016xdc}, RICOCHET~\cite{Billard:2016giu} and NUCLEUS~\cite{Angloher:2019flc}. We also explore the potential for probing these vector mediators through $\nu_e-e^-$ scattering~\cite{Lindner:2018kjo} using a   liquid xenon (LXe)  detector exposed to neutrinos from a $^{51}$Cr~\footnote{This follows the same spirit as the proposal in~\cite{Barabanov:1998bj}. Note also that a proposal for measuring \cevns with a $^{51}$Cr source also exists, see Ref.~\cite{Bellenghi:2019vtc}.} source~\cite{Coloma:2014hka}.

Our paper is organized as follows. In Sec.~\ref{sec:Zprime-contribution-to-cevns} 
we introduce the formalism for \cevns and neutrino-electron scattering in left-right and $E_6$ theories.
The various experimental setups using both SNS and reactor neutrinos for the case of \cevns, as well as $^{51}$Cr neutrinos for the case of neutrino-electron scattering  are described in 
Sec.~\ref{sec:experimental-setup}.
Finally, in Sec.~\ref{sec:results} we present our numerical results for the expected sensitivities to the  mass of  $Z^\prime$ gauge bosons  in the context of the models discussed.


\section{\cevns and neutrino-electron scattering within $Z^\prime$ models}
\label{sec:Zprime-contribution-to-cevns}

In this section we first introduce the notation relevant for the description of \cevns and $\nu_e-e^-$ cross sections in the SM.
We provide the new couplings in the neutrino-quark and neutrino-charged lepton sectors present in extended electroweak models based on left-right or $E_6$ gauge symmetries.
Next, we discuss their subleading effect on the dominant \sm cross sections.

We start from the neutral-current interaction cross section of a neutrino with energy $E_\nu$ scattering off a nucleus with $Z$
protons, $N=A-Z$ neutrons ($A$ is the mass number) and mass $m_A$. In the framework of \sm interactions only and for sufficiently low
momentum transfer, the \cevns channel dominates the cross section, provided that the coherence condition $q \leq 1/R$ ($R$ is the nuclear
radius) is satisfied. Assuming a four-fermion contact interaction, the relevant \cevns cross section can be expressed in terms of the nuclear recoil energy $T_A$ as~\cite{Barranco:2005yy,Kosmas:2017tsq}

\begin{equation}
\left(\frac{d \sigma}{dT_A}\right)_{\text{SM}} = \frac{G_F^2 m_A}{\pi}  \mathcal{Q}_V^2 \left(1 - \frac{m_A T_A}{2 E_\nu^2}\right)  F^2(Q^2) \, ,
\label{eq:xsec-cevns}
\end{equation}
where $G_F$ denotes the Fermi constant and $\mathcal{Q}_V$ is the vector weak charge written in the form
\begin{equation}
\mathcal{Q}_V =  \left[ 2(g_{u}^{L} + g_{u}^{R}) + (g_{d}^{L} + g_{d}^{R}) \right] Z  + \left[ (g_{u}^{L} + g_{u}^{R}) +2(g_{d}^{L} + g_{d}^{R}) \right] N   \, .
\label{eq:CEvNS_couplings}
\end{equation}
For later convenience, $\mathcal{Q}_V$ is expressed in terms of the left- and right-handed couplings of the quark $q=\{u,d\}$ to the $Z$-boson, as
\begin{equation}
\begin{aligned}
g_{u}^{L} =& \rho_{\nu N}^{NC} \left( \frac{1}{2}-\frac{2}{3} \hat{\kappa}_{\nu N} \hat{s}^2_Z \right) + \lambda_u^L \, ,\\
g_{d}^{L} =& \rho_{\nu N}^{NC} \left( -\frac{1}{2}+\frac{1}{3} \hat{\kappa}_{\nu N} \hat{s}^2_Z \right) + \lambda_d^L  \, ,\\
g_{u}^{R} =& \rho_{\nu N}^{NC} \left(-\frac{2}{3} \hat{\kappa}_{\nu N} \hat{s}^2_Z \right) + \lambda_u^R  \, ,\\
g_{d}^{R} =& \rho_{\nu N}^{NC} \left(\frac{1}{3} \hat{\kappa}_{\nu N} \hat{s}^2_Z \right) + \lambda_d^R \, , 
\end{aligned}
\end{equation}
with the weak mixing angle taken in the $\footnotesize{\overline{\text{MS}}}$ scheme, i.e.  $\hat{s}^2_Z = 0.2312$. The radiative corrections from the 
Particle Data Group:  $\rho_{\nu N}^{NC} = 1.0082$, $\hat{\kappa}_{\nu N} = 0.9972$, $\lambda_u^L = -0.0031$, $\lambda_d^L = -0.0025$ and 
$\lambda_d^R =2\lambda_u^R = 3.7 \times 10^{-5}$ are also included. 

In the present study, important corrections due to the finite nuclear size are incorporated through the momentum variation of the nuclear form factors $F(Q^2)$.
These lead to a suppression of the expected \cevns event rate. A comprehensive analysis of the form factor effects has been recently conducted in 
Refs.~\cite{Cadeddu:2017etk,Papoulias:2019lfi} using the first COHERENT data. Here we consider the symmetrized Fermi (SF) approximation~\cite{Sprung_1997} 
\begin{equation}
F \left( Q ^ { 2 } \right) =  \frac { 3 } { Q c \left[ ( Q c ) ^ { 2 } + ( \pi Q a ) ^ { 2 } \right] }  \left[ \frac { \pi Q a } { \sinh ( \pi Q a ) } \right]   \left[ \frac { \pi Q a \sin ( Q c ) } { \tanh ( \pi Q a ) } - Q c \cos ( Q c ) \right] \, ,
\end{equation}
with 
\begin{equation}
c = 1.23 A^{1/3} - 0.60 \, \text{(fm)}, \quad a=0.52 \, \text{(fm)} \, ,
\label{eq:SF-vals}
\end{equation}
where $c$ and $a$ represent the half-density radius and diffuseness, respectively.

Within the SM, the differential cross section describing $\nu_e-e^-$  scattering arises from both neutral- and charged-current interactions and reads~\cite{Garces:2011aa}
\begin{equation}
\frac{d \sigma}{d T_e} (E_\nu, T_e)=\frac{2 G_{F} m_{e}}{\pi}\left[(g_e^{L})^{2}+(g_e^{R})^{2}\left(1-\frac{T_e}{E_{\nu}}\right)^{2}-g_e^{L} g_e^{R} \, \frac{m_{e} T_e}{E_{\nu}^{2}}\right] \, ,
\label{eq:exsec-nue}
\end{equation}
where the QED corrections have been neglected and the chiral couplings take the form 
\begin{equation}
\begin{aligned} 
g_e^{L} &=\rho_{\nu e} \left(-\frac{1}{2} + \hat{\kappa}_{\nu e} \, \hat{s}_Z^{2} \right)  + 1\,  ,\\
g_e^{R} &=\rho_{\nu e} \, \hat{\kappa}_{\nu e} \, \hat{s}_Z^{2}  \, ,
\end{aligned}
\label{eq:nu-e_couplings}
\end{equation}
with the radiative corrections $\rho_{\nu e}=1.0128$ and $\hat{\kappa}_{\nu e}=0.9963$.

We now proceed with our discussion by expressing the new couplings $f_q^{L,R}$ and $f_e^{L,R}$ relevant to \cevns and $\nu_e-e^-$
scattering in a more convenient form. In the context of the $E_6$ and LR symmetric models discussed below, the corresponding beyond the \sm cross sections are obtained through the
substitutions $g_q^{L,R} \to f_q^{L,R}$ and $g_e^{L,R} \to f_e^{L,R}$ in  the cross sections expressions in Eqs.~(\ref{eq:xsec-cevns})~and~(\ref{eq:exsec-nue}) for \cevns and $\nu_e-e^-$ scattering, respectively.

\subsection{Left-right symmetry}

There are various left-right-symmetric models using the gauge group $SU(2)_L\otimes SU(2)_R\otimes U(1)_{B-L}$, restoring the parity symmetry at high energies~\cite{Mohapatra:1980yp}.
These models give an interesting phenomenology, associated to the existence of additional charged and neutral gauge bosons~\cite{GonzalezGarcia:1990yq,GonzalezGarcia:1990tr,Langacker:2008yv,Das:2012ii,Deppisch:2013cya}. 
Here we consider models where the $Z^\prime$ arises from left-right symmetrical extensions of the SM.
In contrast to the charged intermediate vector bosons, it has been shown that the neutral one, $Z^\prime$, can have masses at the TeV scale
consistent with neutrino mass generation and gauge coupling unification in $SO(10)$~\cite{Malinsky:2005bi}.
In what follows, we will focus on the phenomenology coming from such $Z'$ boson. 

\subsubsection{\cevns}

\noindent In the framework of the left-right symmetric model, the relevant parameters describing \cevns are modified as follows~\cite{Polak:1991pc}
\begin{equation}
\begin{aligned}
f^{L}_u=& \rho_{\nu N}^{N C} \mathcal{A} \left(\frac{1}{2}-\frac{2}{3} \hat{\kappa}_{\nu N} \hat{s}_{Z}^{2}\right)- \mathcal{B} \frac{2}{3} \hat{s}_{Z}^{2}+\lambda^{L}_u \, , \\ 
f^{L}_d=&\rho_{\nu N}^{N C} \mathcal{A} \left(-\frac{1}{2}+\frac{1}{3} \hat{\kappa}_{\nu N} \hat{s}_{Z}^{2}\right) + \mathcal{B} \frac{1}{3} \hat{s}_{Z}^{2}+\lambda^{L}_d \, ,\\
f^{R}_u=&\rho_{\nu N}^{N C} \mathcal{A} \left(-\frac{2}{3} \hat{\kappa}_{\nu N} \hat{s}_{Z}^{2}\right) + \mathcal{B} \left(\frac{1}{2}-\frac{2}{3} \hat{s}_{Z}^{2}\right)+\lambda^{R}_u \, , \\ 
f^{R}_d=& \rho_{\nu N}^{N C} \mathcal{A} \left(\frac{1}{3} \hat{\kappa}_{\nu N} \hat{s}_{Z}^{2}\right) + \mathcal{B} \left(-\frac{1}{2}+\frac{1}{3} \hat{s}_{Z}^{2}\right)+\lambda^{R}_d \, ,
\end{aligned}
\end{equation}
with the definitions
\begin{equation}
\mathcal{A} = 1+\frac{\hat{s}_{Z}^{4}}{1-2 \hat{s}_{Z}^{2}} \gamma \, , \qquad \mathcal{B} =\frac{\hat{s}_{Z}^{2} \left(1-\hat{s}_{Z}^{2}\right)}{1-2 \hat{s}_{Z}^{2}}\gamma \, ,
\label{eq:AB}
\end{equation}
and  $\gamma=\left(M_{Z} / M_{Z^{\prime}}\right)^{2}$, where $M_{Z^{\prime}}$ denotes the $Z^\prime$ mass.
\subsubsection{Neutrino-electron scattering}

\noindent Turning to the case of neutrino-electron scattering in the left-right symmetric model, the relevant couplings are trivially obtained as
\begin{equation}
\begin{aligned}
f_{e}^{L}= & \mathcal{A} g_e^{L} + \mathcal{B} g_e^{R} \, , \\
f_{e}^{R}=& \mathcal{A} g_e^{R} + \mathcal{B} g_e^{L} \, ,
\end{aligned}
\end{equation}
where the dependence on the $Z^\prime$ mass is incorporated through the parameters $\mathcal{A}$ and $\mathcal{B}$ defined as in the case of CE$\nu$NS in Eq.~(\ref{eq:AB}).

\subsection{$E_6$ models}
\label{sec:E6}

New neutral gauge bosons also appear in the primordial $E_6$ gauge symmetry~\cite{GonzalezGarcia:1990yq,GonzalezGarcia:1990tr,Langacker:2008yv}.
Since it is a rank-six group, $E_6$ in general yields two neutral gauge bosons beyond those present in the SM. 
These gauge bosons couple to two new hypercharges, $\chi$ and $\psi$ that correspond to the $U(1)$ symmetries present in $E_6/SO(10)$ 
and in $SO(10)/SU(5)$. The corresponding hypercharge quantum numbers are given in Table~\ref{tab:E6}.
We assume that, at low energies, there is only one $U(1)$ symmetry, written as the combination of the symmetries $U(1)_\chi$ and $U(1)_\psi$.  
This defines a one-parameter family of models with hypercharge given  as 
\begin{equation}
Y_\beta =Y_{\chi}\cos \beta +Y_{\psi}\sin \beta \, , 
\end{equation}
whereas the charge operator takes the usual form  $ Q=T^3+ Y $. 
Within this framework, we can write the expressions for the low-energy effective Lagrangian and compute the corresponding corrections to the
SM couplings. 
\begin{table}
   
\begin{tabular}{|c|c|c|c|}
\hline  & $T_3$ & $\sqrt{40}Y_\chi$ &
$\sqrt{24}Y_\psi$   \\
\hline
$Q$   & $\begin{pmatrix} 1/2 \\ -1/2 \end{pmatrix}$ & $-1$ & $1$ \\
$u^c$ &       $0$               & $-1$ & $1$ \\
$e^c$ &       $0$               & $-1$ & $1$ \\
$d^c$ &       $0$               & $ 3$ & $1$ \\
$l$   & $ \begin{pmatrix} 1/2 \\ -1/2 \end{pmatrix}$ & $ 3$ & $1$ \\
\hline
\end{tabular}  
\caption{Hypercharge quantum numbers for the \sm fermions under $E_6$.}
\label{tab:E6}
\end{table}

\subsubsection{Coherent elastic neutrino-nucleus scattering}

\noindent In the context of the $E_6$ model, the new couplings read
\begin{equation}
\begin{aligned}
f_{u}^{L} =& g_{u}^{L} + \varepsilon_u^L \, ,\\
f_{d}^{L} =& g_{d}^{L} + \varepsilon_d^L \, ,\\
f_{u}^{R} =& g_{u}^{R} + \varepsilon_u^R \, ,\\
f_{d}^{R} =& g_{d}^{R} + \varepsilon_d^R \, , 
\end{aligned}
\end{equation}
where the $\varepsilon_{q}^{P}$ contributions are written as~\cite{Barranco:2007tz}
\begin{equation}
\begin{aligned}
\varepsilon_u^L=&-4 \gamma \hat{s}_Z^{2} \rho_{\nu N}^{N C}\left(\frac{c_{\beta}}{\sqrt{24}}-\frac{s_{\beta}}{3} \sqrt{\frac{5}{8}}\right)\left(\frac{3 c_{\beta}}{2 \sqrt{24}}+\frac{s_{\beta}}{6} \sqrt{\frac{5}{8}}\right) \, , \\
\varepsilon_d^R=&-8 \gamma \hat{s}_Z^{2} \rho_{\nu N}^{N C}\left(\frac{3 c_{\beta}}{2 \sqrt{24}}+\frac{s_{\beta}}{6} \sqrt{\frac{5}{8}}\right)^{2} \, , 
\\
\varepsilon_d^L=& \varepsilon_u^L=-\varepsilon_u^R \, ,
\end{aligned}
\label{eq:E6-couplings}
\end{equation}
with the abbreviations $c_{\beta}=\cos \beta$ and  $s_{\beta}=\sin \beta$. 
Three different $E_6$ models are considered here, namely the $(\chi, \psi, \eta )$ models corresponding to $\cos \beta=(1, 0, \sqrt{3/8})$. 
Note that, for $\cos \beta =(-\sqrt{5/32}, 0)$, the new physics contributions vanish, and therefore, there is no sensitivity to $Z^\prime$, i.e.
the $\psi$ model can not be probed in \cevns studies.

\subsubsection{Neutrino-electron scattering}

\noindent For this case the relevant couplings read
\begin{equation}
\begin{aligned} 
f^{L}_e & = g^{L}_e + \varepsilon^{L}_e \, ,\\
f^{R}_e & = g^{R}_e + \varepsilon^{R}_e \, ,
\end{aligned}
\end{equation}
with the new contributions written as
\begin{equation}
\begin{aligned}
\varepsilon^{L}_e=& 2 \gamma \hat{s}_Z^{2} \rho_{\nu e} \left(\frac{3 c_{\beta}}{2 \sqrt{6}}+\frac{s_{\beta}}{3} \sqrt{\frac{5}{8}}\right)^{2} \, , \\ 
\varepsilon^{R}_e =& 2 \gamma \hat{s}_Z^{2} \rho_{\nu e} \left(\frac{c_{\beta}}{2 \sqrt{6}}-\frac{s_{\beta}}{3} \sqrt{\frac{5}{8}}\right)\left(\frac{3 c_{\beta}}{\sqrt{24}}+\frac{s_{\beta}}{3} \sqrt{\frac{5}{8}}\right) \, ,
\end{aligned}
\end{equation} 
and $\gamma$  defined as previously in the \cevns case. Here, it is interesting to note that, for $\cos \beta =- \sqrt{5/32}$,
the coupling constants are equal to zero, so there is no sensitivity to new physics in this case~\cite{Barranco:2007tz}.

\section{Experimental setups}
\label{sec:experimental-setup}

We now examine a number of conceivable experimental setups that may be used to probe for the existence of new neutral gauge bosons.
In particular, we consider \cevns experiments employing both SNS and reactor neutrinos with various possible targets, as well as a future $\nu_e-e^-$ scattering experiment using a $^{51}$Cr source.

\subsection{\cevns from accelerator and reactor neutrinos}

For the case of \cevns experiments, the total number of events between the threshold $T_{\mathrm{th}}$ and the maximum nuclear recoil energy allowed by the kinematics, $T_{A}^{\mathrm{max}} = 2 E_\nu^2/m_A$, can be expressed as~\cite{Kosmas:2017tsq} 
\begin{equation}
\begin{aligned}
N_{\mathrm{theor}} = \sum_{\nu_\alpha} \sum_{x = \mathrm{isotope}} \mathcal{F}_x   \int_{T_{\mathrm{th}}}^{T_{A}^{\mathrm{max}}} \int_{\sqrt{m_A T_A /2}}^{E_{\nu}^{\mathrm{max}}} \lambda_{\nu_{\alpha}}(E_\nu) \mathcal{E}( T_A ) \left(\frac{{d \sigma}_{x}}{dT_A}(E_\nu, T_A) \right)_{\mathrm{tot}}   dE_\nu dT_A \, .
\end{aligned}
\label{eq:events}
\end{equation}
As indicated, the sum is taken over the detector isotopes, $x$, and the neutrino flavors, $\alpha$. In this expression, $\mathcal{F}_x = N_{\mathrm{targ}}^x \Phi_\nu$ denotes the corresponding luminosity on
the detector. This depends on the neutrino flux at the detector, $\Phi_\nu (L)$, and the number of target nuclei, $N_{\mathrm{targ}}^x$, (see Table~\ref{table:exper}). 
The efficiency function $\mathcal{E}(T_A)$ for each given experiment is taken according to Table~\ref{table:exper} as well.

For the pion decay at rest ($\pi$-DAR) neutrinos, relevant for the COHERENT experiment, the neutrino energy distributions are
adequately described by the Michel spectrum~\cite{Louis:2009zza}
\begin{equation}
\begin{aligned} 
\lambda_{\nu_\mu}(E_\nu) & =  \delta\left(E_\nu-\frac{m_{\pi}^{2}-m_{\mu}^{2}}{2 m_{\pi}}\right) \, , \\ 
\lambda_{\bar{\nu}_\mu}(E_\nu) & = \frac{64 E^{2}_\nu}{m_{\mu}^{3}}\left(\frac{3}{4}-\frac{E_\nu}{m_{\mu}}\right) \, ,\\ 
\lambda_{\nu_e}(E_\nu) & = \frac{192 E^{2}_\nu}{m_{\mu}^{3}}\left(\frac{1}{2}-\frac{E_\nu}{m_{\mu}}\right) \, .
\end{aligned}
\label{labor-nu}
\end{equation}
For reactor-based neutrino experiments, we consider the corresponding antineutrino energy distribution $\lambda_{\bar{\nu}_e}(E_\nu)$ resulting from the fission products~\footnote{Note that for the noncommercial reactor used by MINER, only the $^{235}$U contribution is considered.} of $^{235}$U, $^{238}$U, $^{239}$Pu and $^{241}$Pu~\cite{Mention:2011rk}, while for $E_{\bar{\nu}_e} < 2~\mathrm{MeV}$ we rely on the theoretical spectrum given in Ref.~\cite{Kopeikin:1997ve}.

\begin{table*}[t]
\resizebox{\textwidth}{!}{\begin{tabular}{|lcccccc|}
\hline
\textbf{Experiment}  & \textbf{detector} & \textbf{mass} & \textbf{threshold} & \textbf{efficiency}& \textbf{exposure} & \textbf{baseline (m)} \\ 
\hline
 
\multicolumn{7}{|c|}{\textbf{SNS}}\\

COHERENT~\cite{Akimov:2017ade}  & CsI[Na]  & 14.57~kg   &  5~keV    & \cite{Akimov:2018vzs}   & 308.1~days   & 19.3  \\
COHERENT~\cite{Akimov:2018ghi}  & HPGe     & 15~kg      &  5~keV    & 50\%                    &       1~yr   & 22    \\
COHERENT~\cite{Akimov:2018ghi}  & LAr      & 1~ton      & 20~keV   & \cite{Akimov:2019rhz}                    &       1~yr   & 29    \\
COHERENT~\cite{Akimov:2018ghi}  & NaI[Tl]  & 2~ton      & 13~keV    & 50\%                    &       1~yr   & 28    \\
                                
\multicolumn{7}{|c|}{\textbf{Reactor Experiments}}\\

CONUS~\cite{Hakenmuller:2019ecb}                      & Ge         & 3.85~kg   & 100 eV    & 50\%  & 1~yr  & 17   \\
CONNIE~\cite{Aguilar-Arevalo:2019jlr}   & Si         & 1~kg       & ~28 eV    & 50\%     & 1~yr  & 30   \\
MINER~\cite{Agnolet:2016zir}            & 2Ge:1Si    & 1~kg       & 100 eV    & 50\%   & 1~yr  & ~2   \\
TEXONO~\cite{Wong:2010zzc}              & Ge         & 1~kg       & 100 eV    & 50\%   & 1~yr  & 28   \\
RED100~\cite{Akimov:2016xdc}            & Xe         & 100~kg     & 500 eV    & 50\%   & 1~yr  & 19   \\ 

RICOCHET~\cite{Billard:2016giu} & (Zn, Ge) & (1~kg, 1~kg) & (50~eV, 50~eV)  & 50\% & 1~yr  & 100 \\
NUCLEUS~\cite{Angloher:2019flc} &  \begin{tabular}{cc} ($\mathrm{CaWO_4}$, & $\mathrm{Al_2 O_3}$)\\
 (Ge, & Si)
\end{tabular} & \begin{tabular}{cc} (4.41~gr, & 6.84~gr) \\ (0.5~kg, & 0.5~kg) \end{tabular}  & \begin{tabular}{cc} 20~eV \\ 50~eV \end{tabular} & 50\% & 1~yr  & 100 \\
\hline
\end{tabular}}
\caption{CE$\nu$NS experiments and various setups considered in the present study.} 
\label{table:exper}
\end{table*}
%

\subsection{Neutrino-electron scattering from a $\mathrm{^{51}Cr}$ source}

Another experimental configuration that we have studied uses neutrinos from an artificial neutrino source, as suggested in~\cite{Barabanov:1998bj}~\footnote{For a study focusing on $\bar{\nu}_e-e^-$ scattering at the TEXONO experiment see Ref.~\cite{Deniz:2017zok}. }.
Following Ref.~\cite{Coloma:2014hka}, we consider for this case a cylindrical LXe detector with height $h=1.38$~m, diameter $d=1.38$~m, located at $L = 1$~m above
a 1 MCi radioactive $^{51}$Cr source with a flux $\phi_0= 2.94 \times 10^{15}$ $\mathrm{\nu/(MCi~m^2~s)}$. The emitted neutrino spectra consist of
  two monochromatic beams with energies $E_{\nu_1}=430$~keV and $E_{\nu_2}=750$~keV with a relative strength $\alpha_1=10\%$ and $\alpha_2=90\%$, respectively.
Due to the exponentially decaying nature of the source within a time interval $\Delta t$,  we take the time-averaged activity~\cite{Coloma:2014hka},
\begin{equation}
\left\langle R_{\mathrm{Cr} 51}\right\rangle =\frac{\tau R_{\mathrm{Cr} 51}^{0}}{\Delta t}\left[1-e^{-\Delta t / \tau}\right] \, ,
\end{equation}
where $R_{\mathrm{Cr} 51}^{0}$ denotes the initial radioactivity and $\tau=39.96$~days is the mean lifetime of $^{51}$Cr. The number of neutrino-electron scattering events  within a bin $i$ with recoil energy in the range $[T_{e,i}, T_{e,i}+\delta T_e]$ and maximum recoil energy $T_e^{\mathrm{max}} = 2 E_\nu^2 / (2 E_\nu + m_e)$ is given by
\begin{equation}
N^i_{\mathrm{events}} = \mathcal{F} \int_{T_{e,i}}^{T_{e,i}+\delta T_e} \left[ \alpha_1\frac{d \sigma}{d T_e} (E_{\nu_1}, T_e) + \alpha_2\frac{d \sigma}{d T_e} (E_{\nu_2}, T_e) \right] dT_e \, ,
\end{equation}
with $\mathcal{F} = \Phi_{\text{avg}}^{\mathrm{^{51}Cr}} V n_e \Delta t$. Here, $V$ represents the detector fiducial volume, $n_e$ is the electron density of the target material, while the factor 
$
\Phi_{\text{avg}}^{\mathrm{^{51}Cr}} = \phi_0 \frac{1 \mathrm{m^2}}{r^2_{\mathrm{avg}}} \left\langle R_{\mathrm{Cr} 51}\right\rangle
$
represents the average neutrino flux.
  Finally, due to its cylindrical geometry, the average distance $r_{\mathrm{avg}}$ between the source and the detector is written as~\cite{Link:2019pbm}
\begin{equation}
\begin{aligned}
r_{\mathrm{avg}}=\Biggl[\frac{4 L}{d^{2} h} \log \frac{L^{2}\left(d^{2}+4(h+L)^{2}\right)}{\left(d^{2}+4 L^{2}\right)(h+L)^{2}} + & \frac{4}{d^{2}} \log \frac{d^{2}+4(h+L)^{2}}{4(h+L)^{2}} \\
- & \frac{4}{d h} \tan ^{-1}\left(\frac{2 L}{d }\right)+\frac{4}{d h} \tan ^{-1}\left(\frac{2(h+L)}{d}\right) \Biggr]^{-1/2} \, .
\end{aligned}
\end{equation}
As a test case, in our calculations we assume the three configurations assumed in Ref.~\cite{Link:2019pbm}, (A, B, C): with $R_{\mathrm{Cr} 51}^{0}= (5, 5, 10)$  MCi 
$^{51}$Cr and a time interval $\Delta t= (100, 50, 50)$ days, respectively.

\section{Numerical Results}
\label{sec:results}

We now perform a statistical analysis of the different experimental configurations discussed in the previous sections. 
Our present study is based on a $\chi^2$ fit of the measured (COHERENT with CsI detector) or expected (other \cevns experiments) number of events. 
We minimize over the nuisance parameters and probe the $Z^\prime$ mass by computing 
$\Delta \chi^2(\mathcal{S})= \chi^2(\mathcal{S}) - \chi^2_{\mathrm{min}}(\mathcal{S})$ with $\mathcal{S}\equiv \{M_{Z^\prime}, \beta \}$. 
Our statistical analysis of the COHERENT data relies on the $\chi^2$ function~\cite{Akimov:2017ade} 
\begin{equation}
\begin{aligned}
\chi^2 (\mathcal{S}) =  \underset{\mathtt{a}_1, \mathtt{a}_2}{\mathrm{min}} \Bigg [ \frac{\left(N_{\mathrm{meas}} - N_{\mathrm{theor}}(\mathcal{S}) [1+\mathtt{a}_1] - B_{0n} [1+\mathtt{a}_2] \right)^2}{(\sigma_{\mathrm{stat}})^2} 
  + \left(\frac{\mathtt{a}_1}{\sigma_{\mathtt{a}_1}} \right)^2 + \left(\frac{\mathtt{a}_2}{\sigma_{\mathtt{a}_2}} \right)^2 \Bigg ] \, ,
\end{aligned}
\label{eq:chi}
\end{equation}
with $N_{\mathrm{meas}}=142$ (measured number of events), $\sigma_{\mathtt{a}_1} = 0.28$ (normalization uncertainty on the signal events), and  $\sigma_{\mathtt{a}_2} = 0.25$  (normalization uncertainty on the background events). $N_{\mathrm{theor}}$ denotes the calculated number of events in the left-right or $E_6$ model.
The statistical uncertainty is calculated as $\sigma_{\mathrm{stat}}=\sqrt{N_{\mathrm{meas}} + B_{0n} + 2 B_{ss}}$
with $B_{0n}=6$ and $B_{ss}=405$ being the prompt-neutron and steady-state background events respectively, (see Refs.~\cite{Akimov:2017ade,Akimov:2018vzs} for more details). 

For the analysis of future \cevns data expected at the Ge, LAr and NaI detector subsystems at COHERENT, as well as at the different reactor-based  
experiments, we consider a single nuisance parameter $\mathtt{a}$ and assign conservative values for the systematic uncertainty, namely $\sigma_{\text{sys}} = 0.2$.  
Although quite simplified, we think this analysis is justified at present.  The $\chi^2$ function in this case reads
\begin{equation}
\chi^2(\mathcal{S}) = \underset{\mathtt{a}}{\mathrm{min}} \Bigg [ \frac{\left(N_{\text{SM}} - N_{\text{theor}}(\mathcal{S}) [1+\mathtt{a}]\right)^2} {(\sigma_\text{stat})^2}   + \left( \frac{\mathtt{a}}{\sigma_{\text{sys}}}\right)^2 \Bigg] \, ,
\label{eq:chi_future}
\end{equation}
where $N_{\text{SM}}$ represents the number of events assuming purely \sm interactions and, as previously, $N_{\text{theor}}$ is the calculated number of events in the presence of LR and $E_6$ interactions, while the estimated statistical uncertainty is taken to be $\sigma_{\text{stat}}=\sqrt{N_{\text{SM}} + N_{\text{bg}}}$. We assume a flat  steady-state background $N_{\text{bg}} =  \sigma_{\text{bg}} N_{\text{SM}}$, with  $\sigma_{\text{bg}} = 0.2$.
 It should be noted that, for the CONUS experiment, a realistic background level of 1--3 cpd has been previously assumed~\cite{Lindner:2016wff} while, for the case of Ricochet, a likelihood analysis with a binned uncertainty following a Poisson distribution has been considered in Ref.~\cite{Billard:2016giu}. For other reactor-based experiments, background-related information is given in Ref.~\cite{Akimov:2019wtg}.

\begin{figure}[t]
\includegraphics[width=\textwidth]{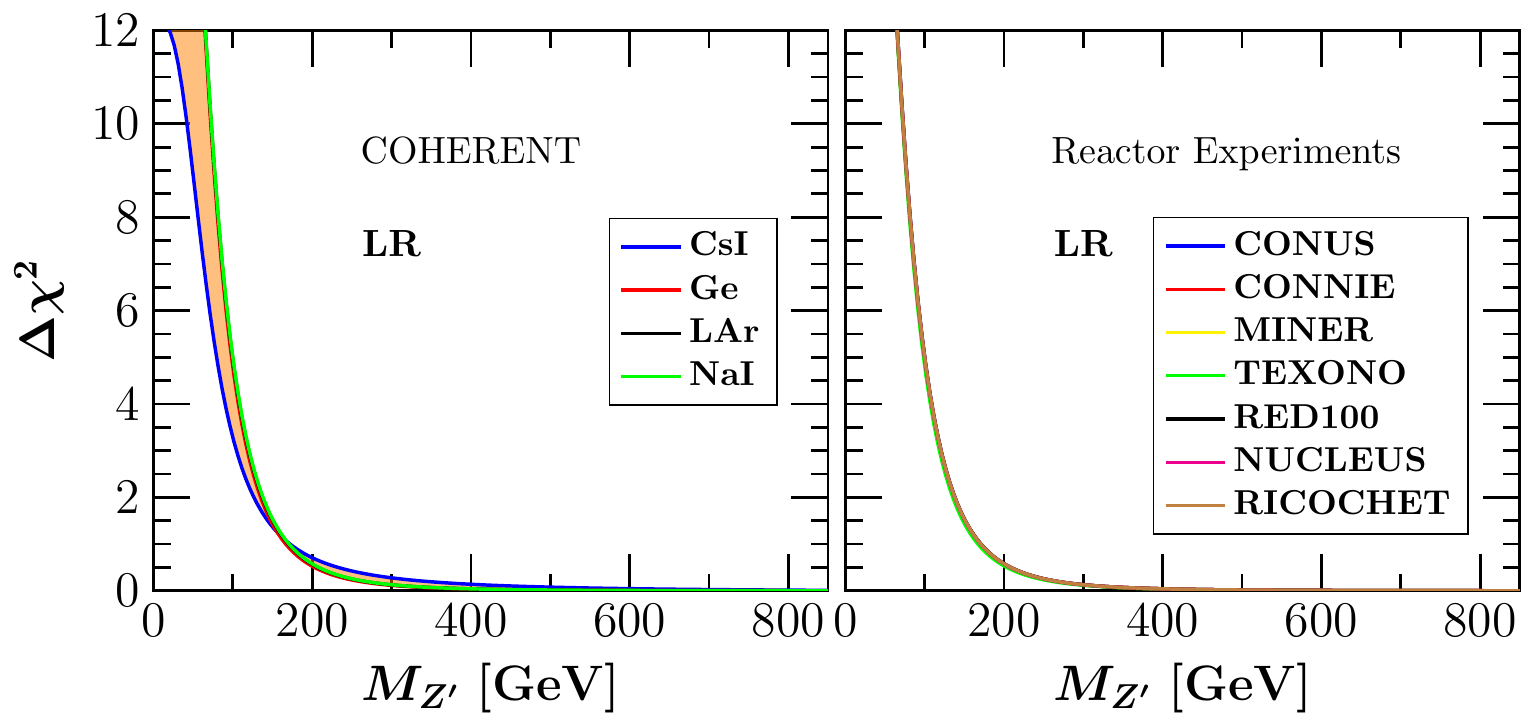}
\includegraphics[width=\textwidth]{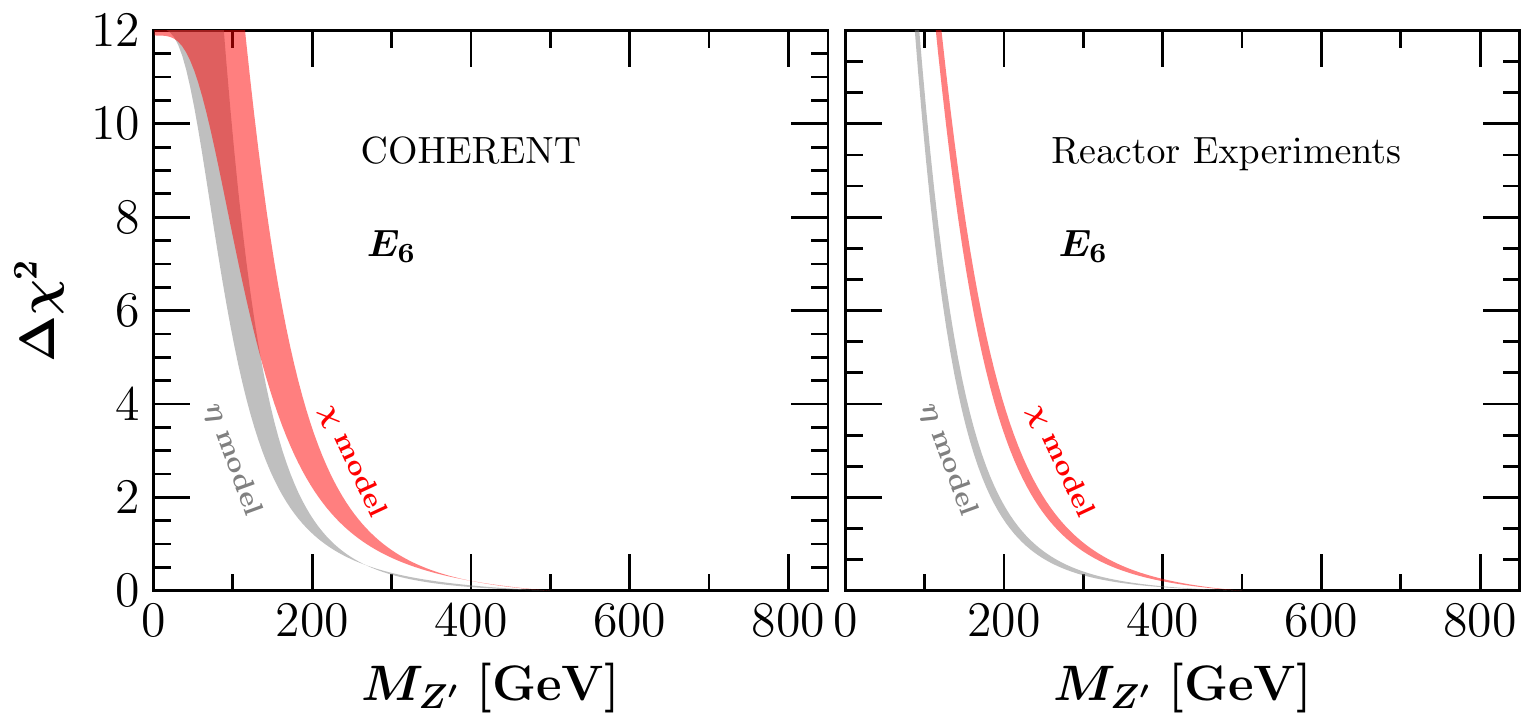}
\caption{$\Delta \chi^2$ profiles for the $Z^\prime$ mass in the left-right-symmetric model (upper panel)  and  $E_6$ models (lower panel).
  The results correspond to \cevns at the SNS (left) and at reactor-based experiments (right).
  For the case of $E_6$ models the red (gray) band corresponds to the sensitivity on $M_{Z^\prime}$ in the $\chi$ ($\eta$) model.}
\label{fig:chi2_LR_E6}
\end{figure}

As a first step,  we obtain the sensitivity on $M_{Z^\prime}$ in the framework of the LR-symmetric model from the available data of COHERENT in terms of a $\chi^2$ fit as described above. 
In a similar manner, we estimate the projected sensitivities at the future SNS and reactor experiments looking for \cevns events. 
The results are presented in the upper-left (upper-right) panel of Fig.~\ref{fig:chi2_LR_E6} for the case of SNS (reactor) experiments. 
From this analysis, it becomes evident that for all the setups the sensitivities are rather poor compared to current bounds from the LHC~\cite{Aad:2019fac},
i.e. we find that $M_{Z^\prime} \gtrsim 125$~GeV at 90\% C.L.

We now turn to the $Z^\prime$ models obtained in the context of $E_6$ symmetry. 
In particular, we explore the corresponding sensitivity on $M_{Z^\prime}$ in the ($\chi$, $\psi$, $\eta$) realizations of $E_6$ by fixing 
$\cos \beta=$ (1, 0, $\sqrt{3/8}$). 
For each model we extract the sensitivity on $M_{Z^\prime}$ assuming the available \cevns data as well as the data expected in the future. 
The results obtained are presented in the lower panel of Fig.~\ref{fig:chi2_LR_E6}, where the red (gray) band corresponds to the $\chi$ ($\eta$) model.
Note that, in comparison with the upper plots in Fig.~\ref{fig:chi2_LR_E6}, here the color labels corresponding to each experiment are dropped. The band width illustrates the sensitivity range considering all the experiments: for the COHERENT experiment, the least (most) constraining detector is the CsI (NaI), as can be seen in the upper panel of Fig.~\ref{fig:chi2_LR_E6}, while for reactor-based facilities one finds that essentially all experiments have the same sensitivity. 
In both cases, the sensitivity to the $Z^\prime$ mass at 90\% C.L. is slightly below (above) 200~GeV for the $\eta$ ($\chi$) model.
A summary of the obtained limits at 90\% C.L. is listed in Table~\ref{tab:limits}.
Notice that \cevns is not sensitive to the $\psi$ model, since the $\varepsilon_{p}^{V}=2\varepsilon_{u}^{V}+\varepsilon_{d}^{V}$ and $\varepsilon_{n}^{V}=\varepsilon_{u}^{V}+2\varepsilon_{d}^{V}$
couplings~\footnote{Note that $\varepsilon_q^V = \varepsilon_q^L +\varepsilon_q^R$ with $q=\{u, d\}$.} are vanishing; see Eq.(\ref{eq:E6-couplings}).
Before closing this discussion, we should note that some of our results could, in principle, be extracted by remapping the NSI bounds derived from the current and future \cevns data~\cite{Kosmas:2017tsq,Billard:2018jnl,Miranda:2019skf}. However, the importance of probing for new neutral gauge bosons more than justifies a dedicated study.
\begin{table}[t!]
\resizebox{\textwidth}{!}{%
\begin{tabular}{|l|cccc|ccccccc|c|}
\hline
model  & \multicolumn{4}{c|}{COHERENT} & \multicolumn{7}{c|}{Reactor Experiments}                                  & $\nu_e-e^-$   \\ \hline
       & CsI   & LAr   & Ge    & NaI   & CONUS & CONNIE & MINER & TEXONO & RED100 & NUCLEUS & RICOCHET & $^{51}$Cr-LXe \\
$\chi$ & 183   & 220   & 215   & 217   & 218   & 228    & 223   & 215    & 215    & 222      & 221      & (510, 481, 564) \\
$\eta$ & 144   & 170   & 166   & 168   & 169   & 177    & 173   & 167    & 166    & 172      & 171      & (487, 459, 540) \\
$\psi$  & --   & --   & --  & --  & --   & --   & --   & --    & --    & --      & --      & (224, 211, 250) \\
LR     & 110   & 124   & 122   & 125   & 125  & 125    & 124   & 123    & 125    & 125      & 125      & (483, 455, 534) \\ \hline
\end{tabular}%
}
\caption{Sensitivity at 90\% C.L. on the $Z^\prime$ mass for all experimental configurations assumed in the present study (see Table~\ref{table:exper} and the text). The values of the reported limits are given in GeV units. We also present results for the proposed $^{51}$Cr-LXe experiment, corresponding to the scenarios (A,B,C).}
\label{tab:limits}
\end{table} 
\begin{figure}[t]
\includegraphics[width=\textwidth]{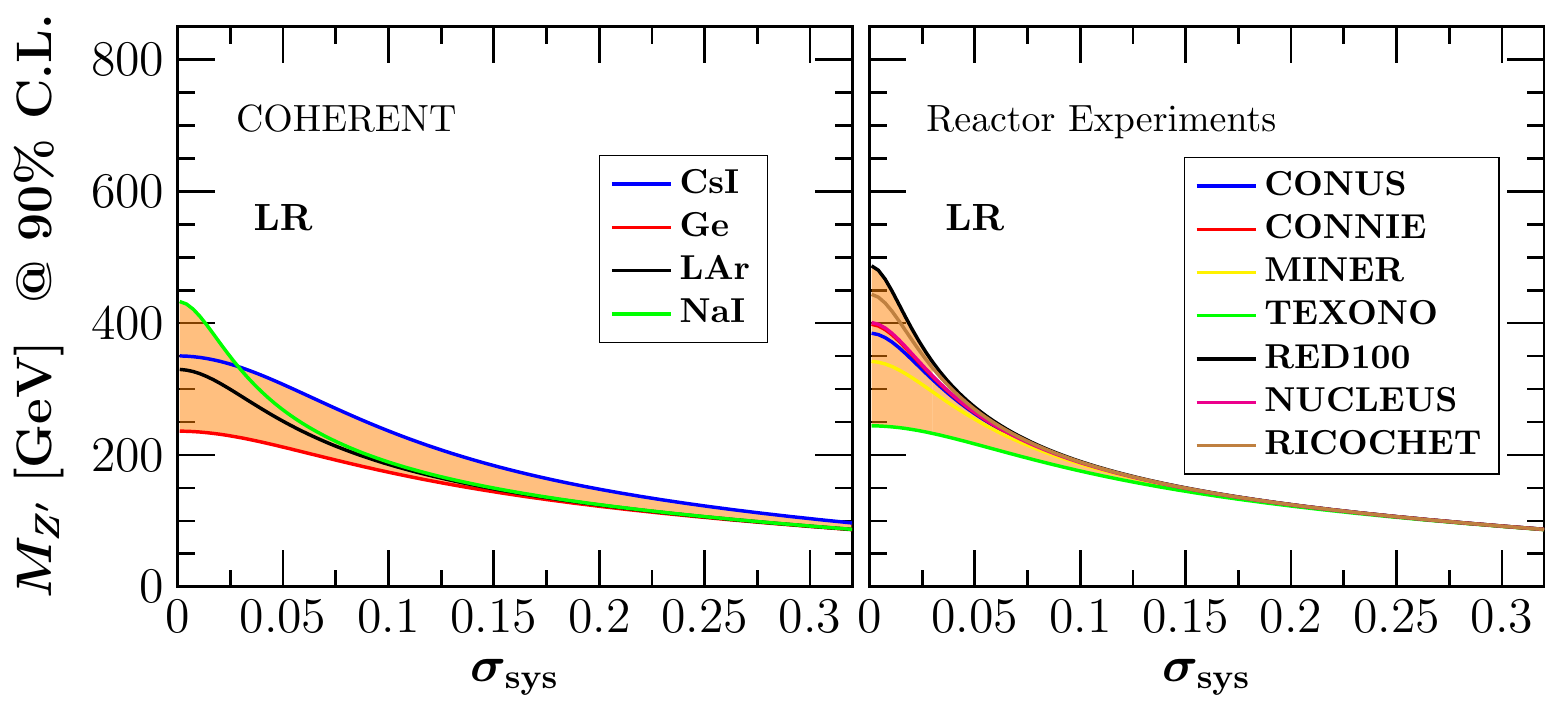}
\includegraphics[width=\textwidth]{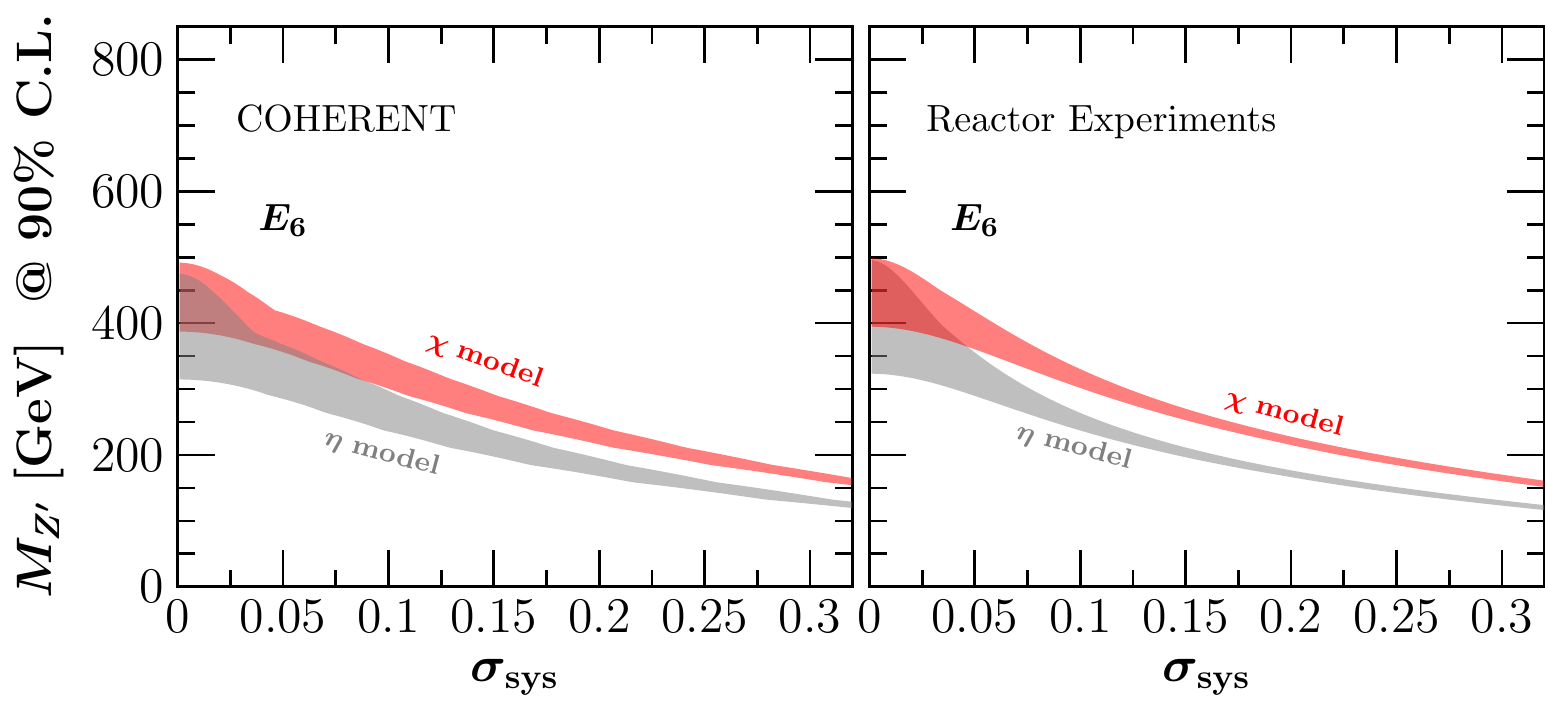}
\caption{ 90\% C.L. sensitivity on the $Z^\prime$ mass as a function of the systematic  uncertainty $\sigma_\text{sys}$. The analysis assumes \cevns experiments at the SNS (left) and at reactor facilities (right). The upper (lower) panel shows the results for the left-right symmetric ($E_6$) model.}
\label{fig:uncertainty}
\end{figure}
\begin{figure}[ht!]
\includegraphics[width=\textwidth]{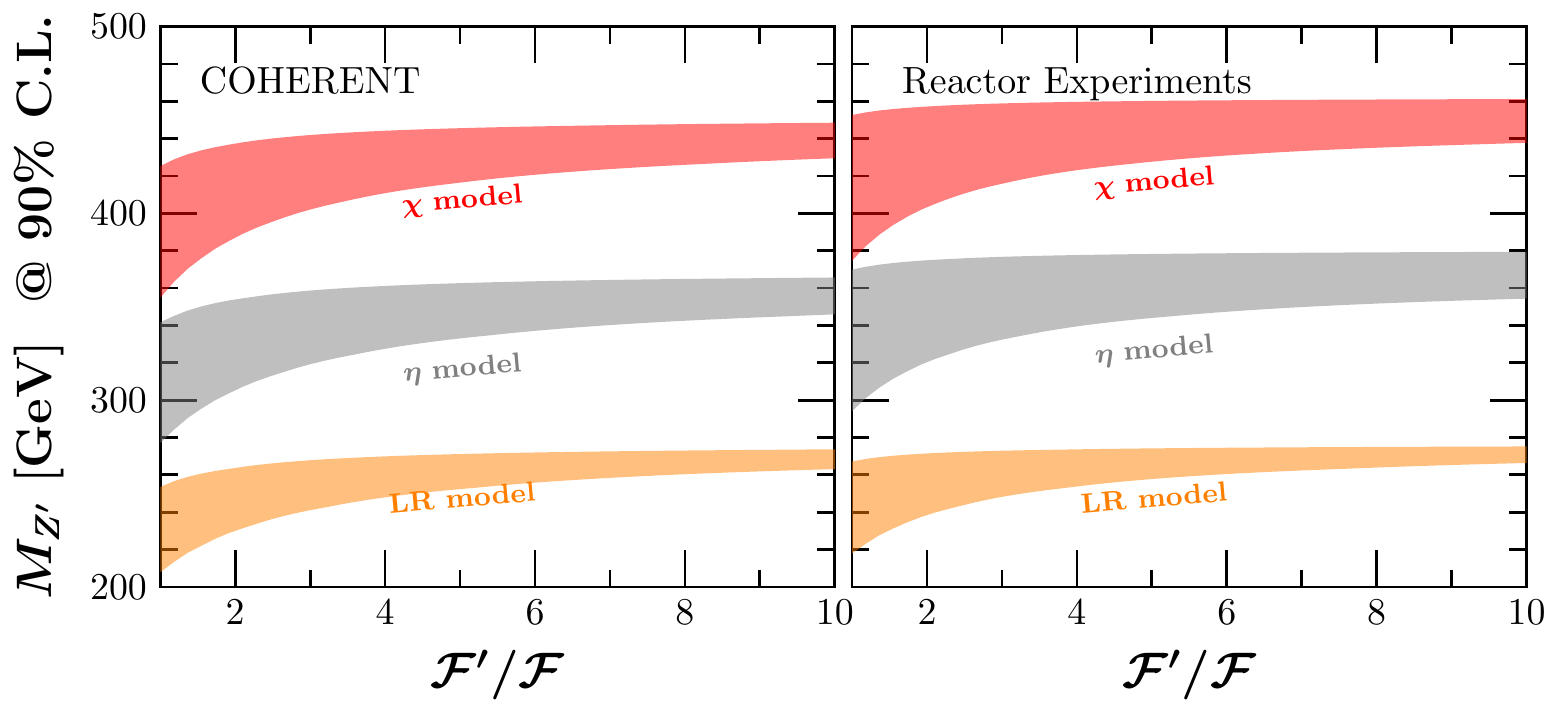}
\caption{90\% C.L. sensitivity on the $Z^\prime$ mass as a function of the total luminosity at a given detector. The results for the $E_6$ and LR symmetric models are shown for the case of \cevns experiments at the SNS (left) and at reactor facilities (right).}
\label{fig:exposure}
\end{figure}

\subsection{Improving the sensitivities on the $Z^\prime$ mass with future \cevns experiments} 

We now explore to what extent the control of uncertainties will offer improved sensitivities on the vector mediator mass. Note that the quenching factor uncertainty is nuclear isotope dependent. For example,  the current uncertainty on CsI is 18.9\%~\cite{Akimov:2017ade}, while for LAr is 2\% ~\cite{Akimov:2019rhz}. Hence, after considering the uncertainties on the flux (10\%), the form factor (5\%) and the acceptance function (5\%), the corresponding total systematic uncertainties for CsI and LAr experiments read 28\% and 12.4\%, respectively. The latter indicates that our earlier assumption of $\sigma_\text{sys}=20\%$ might not be always accurate.
To this purpose, we perform a $\chi^2$ analysis assuming different values for the systematic  uncertainty $\sigma_\text{sys}$, while keeping all other detector specifications fixed according to Table~\ref{table:exper}. 
Reducing the systematic uncertainty is not unreasonable by combining improved quenching factor measurements with a better understanding of the 
nuclear form factors and neutrino fluxes, as well as from the expected substantial improvements on detector technologies aimed at the future 
\cevns experiments. 
Our results are presented in the upper and lower panel of Fig.~\ref{fig:uncertainty} for the LR symmetric and $E_6$ models, respectively. 
As shown previously, the left and right panels show the sensitivities of SNS and reactor \cevns experiments. 
Focusing on the LR symmetric model, it can be seen that, for very low uncertainty, the LAr and NaI detectors perform better while, for larger uncertainty, the
CsI detector is optimal. 
Similarly, for the case of reactor-based \cevns experiments, the xenon-based RED100 (Ge-based TEXONO) appears to have the best (worst) performance. 
The same conclusions are drawn for the case of $E_6$ models where, for convenience, only the bands are displayed.
In both cases, one sees the improvement with respect to Fig.~\ref{fig:chi2_LR_E6}.

At this point, we turn to the impact of neutrino luminosities on improving the attainable sensitivities on the $Z^\prime$ mass at future \cevns experiments. 
We do this by scaling up the number of events, assuming a correspondingly larger detector mass and running period. 
This information is encoded in the future detector luminosity factor, that we denote here as $\mathcal{F}^\prime$. 
We have checked that, with the chosen values of statistical and systematic uncertainties [see Eq.(\ref{eq:chi_future})], the sensitivity shown in Fig.~\ref{fig:chi2_LR_E6} remains practically unaffected by an increase in the exposure. Indeed, the sensitivity on the $Z^\prime$ mass is dominated by the systematic uncertainty at all experiments.
Figure~\ref{fig:exposure} illustrates the projected sensitivity on $M_{Z^\prime}$ at 90\% C.L. as a function of the ratio $\mathcal{F}^\prime/ \mathcal{F}$, 
where $\mathcal{F}$ corresponds to the current or proposed luminosity of each experiment in Table~\ref{table:exper} and  $\sigma_\text{sys} =5\%$. This level of systematic uncertainty can only be reached through a substantial improvement on the quenching factor uncertainty, an improved determination of the nuclear form factors and a better understanding of the neutrino energy distribution.
We therefore conclude that higher intensity \cevns experiments will offer only slightly improved results with respect to those expected from the current experimental setups.
Moreover, one sees that the expected sensitivity for the $\chi$ model is better than that expected in the $\eta$ or LR symmetric models. 

\subsection{Improved $Z^\prime$ sensitivities with future neutrino-electron scattering experiments}

We now expand our analysis by including also information coming from neutrino-electron scattering, which involves both neutral and charged currents.  
As a concrete example, we focus on new interesting proposals that aim to measure neutrino-electron scattering events by employing a LXe detector exposed to neutrino emissions from a radioactive $^{51}$Cr source~\cite{Coloma:2014hka}. Our statistical analysis in this case is based on the $\chi^2$ function
\begin{equation}
\chi^2 = \sum_{i} \left( \frac{N_\text{SM}^i - N_\text{th}^i}{\delta N_\text{SM}^i} \right)^2 \, ,
\end{equation}
with $\delta N^i_\text{SM} = \sqrt{N^i_\text{SM}}$. Since we are dealing with a large number of events, here we have binned the sample with $\delta T_e = \text{5~keV}$, assuming 120 bins in the range $[0, T_e^\text{max}]$. 
\begin{figure}[t]
\includegraphics[width= \textwidth]{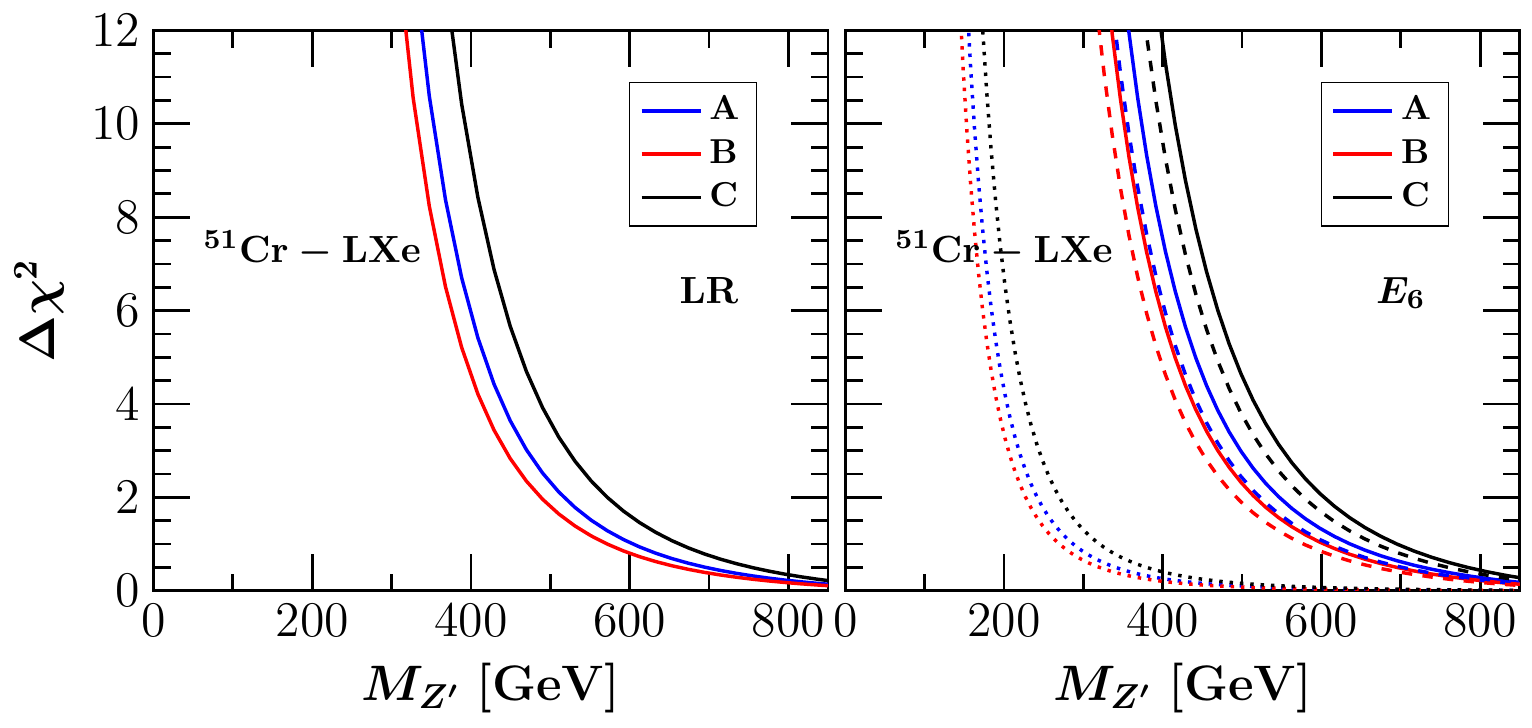}
\caption{Same as Fig.~\ref{fig:chi2_LR_E6} but from the analysis of neutrino-electron scattering using a $^{51}$Cr source. The left panel shows the results for left-right symmetric models. In the right panel, corresponding to $E_6$ models, the solid (dashed) [dotted] curves correspond to the $\chi$ ($\eta$) [$\psi$] model.  The different lines A, B and C correspond to different assumptions for the activity of the source and the exposure, see the text for details.}
\label{fig:chi2_v-e}
\end{figure}
\begin{figure}[ht!]
\includegraphics[width=\textwidth]{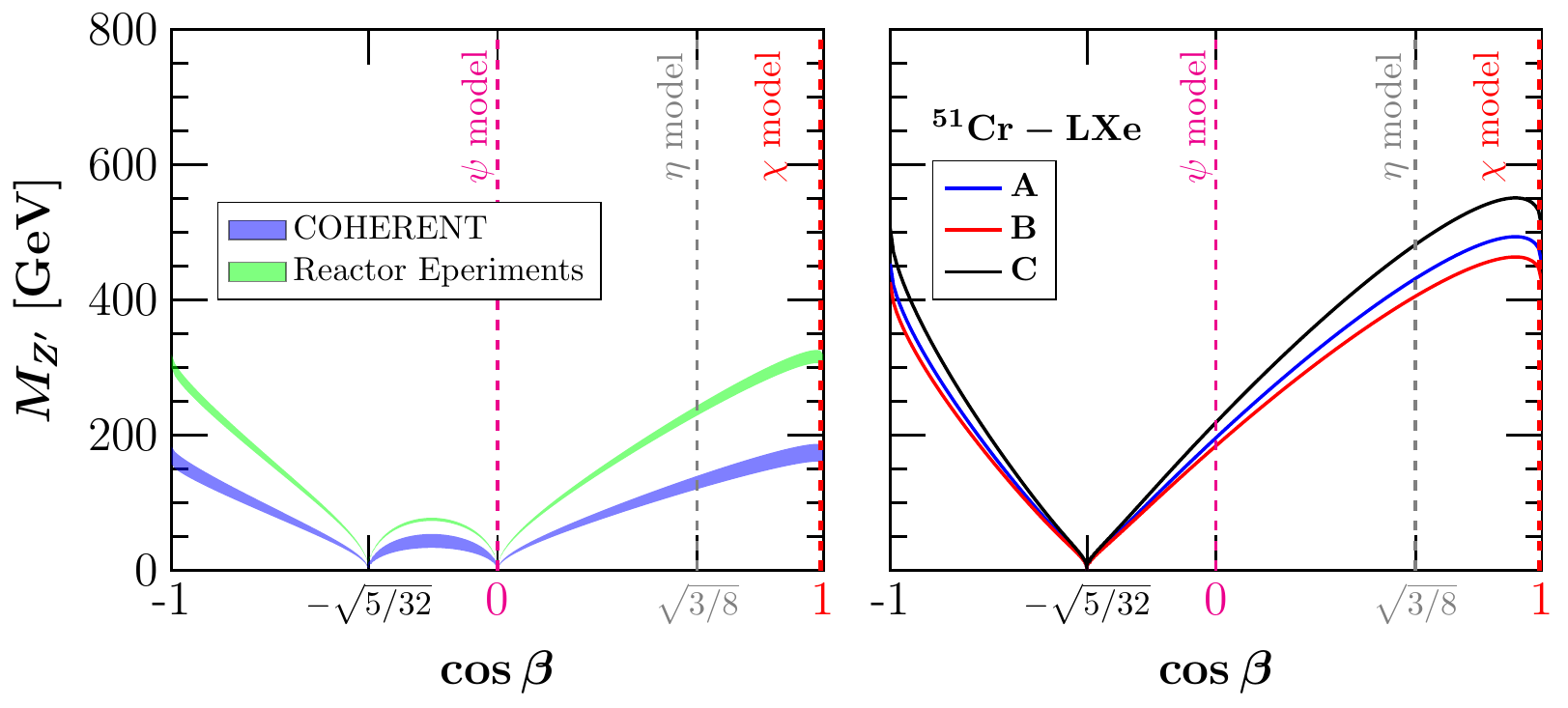}
\caption{90\% C.L allowed regions in the $(\cos \beta$-$M_{Z^\prime})$ plane for \cevns (left) and neutrino-electron scattering (right) experiments. 
}
\label{fig:contours}
\end{figure}
The sensitivities on $M_{Z^\prime}$ for the LR symmetric and $E_6$   models is shown in the left and right panel of Fig.~\ref{fig:chi2_v-e}, respectively. 
As previously, the $\chi$ model (dashed lines) is more sensitive to $M_{Z^\prime}$ compared to the $\eta$ model (solid lines), while the $\psi$ model (dotted lines) is the least sensitive.  
Our results indicate that neutrino-electron scattering within 50-100 days will reach a sensitivity of the order of 200--600~GeV, i.e. 
more competitive with respect to the one extracted from the purely neutral-current CE$\nu$NS. \\ [-.2cm]

We can obtain the sensitivity contours in the $(\cos \beta, M_{Z^\prime})$ plane, where $\beta$ is the parameter defining a one-parameter family of $E_6$ theories, as shown in Fig.~\ref{fig:contours}.
The left panel illustrates the allowed regions obtained from \cevns at the SNS and at reactor experiments, indicated by the shaded blue and green bands, respectively. The right panel shows the corresponding regions from a neutrino-electron scattering experiment using a $^{51}$Cr source, for the different configurations assumed (A, B and C). 
The following conclusions can be extracted from the figure: concerning CE$\nu$NS, the SNS facilities are less sensitive compared to the reactor-based ones, 
while neutrino-electron scattering at LXe with a $^{51}$Cr source is the optimum choice, since it can exclude a larger region of the parameter space. 
Notice as well the presence of two special $\beta$ values for which \cevns experiments have no sensitivity to $M_{Z^\prime}$, since the $Z^\prime$ couplings vanish in this case.
Likewise, one sees that neutrino-electron scattering presents only one such special $\beta$ value with no $Z^\prime$ sensitivity. 
These results are in agreement with our discussion in Sec.~\ref{sec:E6}.

One sees that the potential for probing a new $Z^\prime$ mediator from low-energy measurements of neutrino-electron scattering or \cevns seems to lie well below the sensitivity reached by direct searches at the LHC, i.e. the search for high-mass dilepton resonances produced {\it a la} Drell-Yan~\cite{Aad:2019fac}. 
However, our analysis has been very conservative, as it relies only on the proposed experimental configurations of the first generation of \cevns experiments.
Moreover, we have presented the sensitivity range expected from the various experiments of each type, considering one experiment at a time.

\subsection{Combined expected sensitivities on the $Z^\prime$ mass}

\begin{figure}
\includegraphics[width= 0.8 \textwidth]{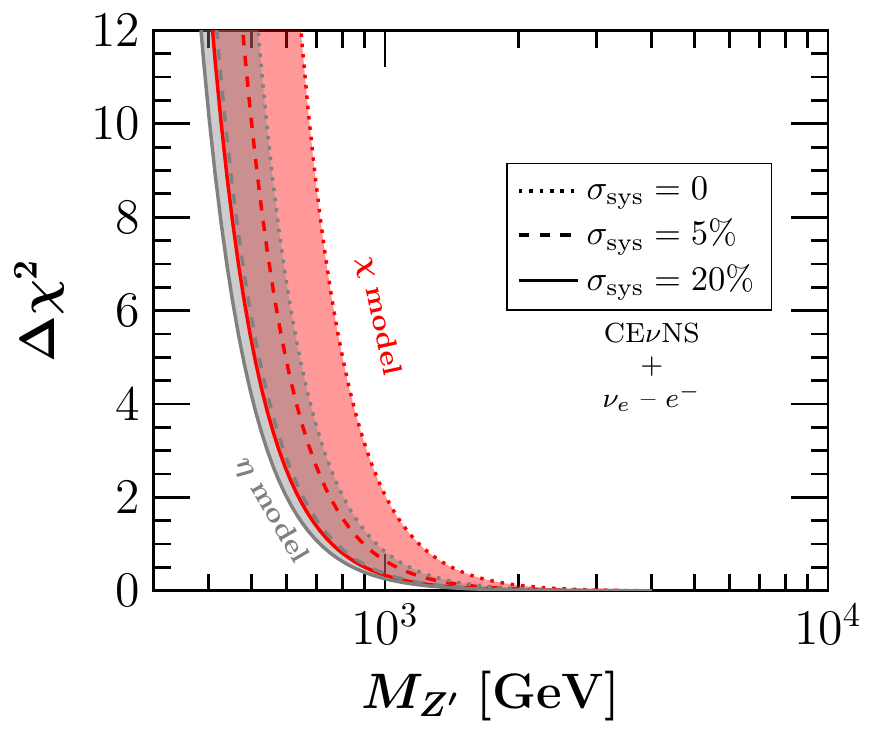}
\caption{Projected sensitivity for the $\chi$ and $\eta$ models of $E_6$ from a future combined analysis of CE$\nu$NS and neutrino-electron scattering data (see text).}
\label{fig:combined_global}
\end{figure}

So far our strategy has been to explore the potential in probing $Z^\prime$ physics within a given low-energy experiment. 
However, one may explore the  phenomenological potential of high intensity low-energy experiments by performing a combined analysis of \cevns and $\nu_e-e^-$ scattering experiments. 
Due to the lack of experimental data, however, only the COHERENT-CsI, CONNIE and $^{51}$Cr-LXe experiments are taken into account. While considering only three experiments for this particular analysis, we however note that different neutrino sources and detector materials are assumed, minimizing the impact of correlation effects of the present analysis.
The corresponding results are illustrated in Fig.~\ref{fig:combined_global} for the $\chi$ and $\eta$ models of $E_6$, assuming different choices of systematic  uncertainties ranging from 20\% to zero. 
Notice that different experimental uncertainties are assumed for the case of \cevns experiments, while for neutrino-electron scattering the C configuration is assumed.
As before, one can also present the sensitivities to the various gauge bosons of $E_6$ models associated to different values of $\beta$. 
One sees that future neutrino-electron scattering and high-intensity \cevns data with a better control of uncertainties have promising prospects for reaching few~TeV scale.
This is the scale currently probed at the high energy frontier experiments, such as the LHC~\footnote{Notice that LHC constraints usually assume the $Z^\prime$ to have SM-strength couplings, and should be adequately rescaled as a function of $\beta$.}. 
It follows that low-energy measurements may offer new probes of $Z^\prime$ parameters, complementary to the high-energy frontier approach.

\section{Conclusions and outlook}

In this work we have quantified the attainable sensitivities on extra neutral gauge bosons $Z^\prime$ that can be reached at high-intensity, low-energy facilities.
We focused on existing and next generation coherent-elastic neutrino-nucleus scattering (CE$\nu$NS) as well as neutrino-electron scattering experiments. 
As neutrino sources, we have discussed the spallation neutron source as well as reactor neutrinos and neutrinos from radioactive sources.
We have concentrated on compelling models that predict the existence of a new neutral vector boson mediator, such as string-inspired $E_6$ schemes and models with left-right symmetry. 
The $Z^\prime$ contributions to the CE$\nu$NS and neutrino-electron scattering in this class of theories were studied.
Current  $Z^\prime$ limits are obtained from Fig.~\ref{fig:chi2_LR_E6} and given in Table~\ref{tab:limits}. 
Future  $Z^\prime$ sensitivities from individual \cevns at SNS and reactor experiments are given in Figs.~\ref{fig:uncertainty} and \ref{fig:exposure}. 
A comparison with the expected sensitivity from a future neutrino-electron scattering using a $^{51}$Cr source and a ton-scale Liquid Xenon detector is also given in Fig.~\ref{fig:chi2_v-e}. 
Expected sensitivities for an arbitrary $E_6$ model are presented in Fig.~\ref{fig:contours}. 
Finally, combined global sensitivities were presented in Fig.~\ref{fig:combined_global}.
The high-intensity, low-energy approach is not only complementary to the high-energy frontier measurements at colliders, but could also become competitive in the long run, as shown in  Fig.~\ref{fig:combined_global}.


\begin{acknowledgments}

This work is supported by the Spanish grants FPA2017-85216-P (AEI/FEDER, UE), PROMETEO/2018/165 (Generalitat
Valenciana) and the Spanish Red Consolider MultiDark FPA2017-90566-REDC. OGM has been supported by CONACYT-Mexico under
grant A1-S-23238 and by SNI (Sistema Nacional de Investigadores). MT acknowledges financial support from MINECO through the Ram\'{o}n y Cajal contract RYC-2013-12438.

\end{acknowledgments}


%
\bibliographystyle{utphys}
\bibliography{bibliography}

\end{document}